\documentclass[%
 aip,
 amsmath,amssymb,
 reprint,%
]{revtex4-1}

\usepackage{graphicx}% Include figure files
\usepackage{dcolumn}% Align table columns on decimal point
\usepackage{bm}% bold math
\usepackage{float}
\usepackage[utf8]{inputenc}
\usepackage[T1]{fontenc}
\usepackage{mathptmx}
\usepackage{etoolbox}
\usepackage{graphicx}
\usepackage{subfig} 
\makeatletter
\def\@email#1#2{%
 \endgroup
 \patchcmd{\titleblock@produce}
  {\frontmatter@RRAPformat}
  {\frontmatter@RRAPformat{\produce@RRAP{*#1\href{mailto:#2}{#2}}}\frontmatter@RRAPformat}
  {}{}
}%
\makeatother
\begin{document}

\preprint{AIP/123-QED}

\title[Sample title]{Experiment Research on Feasibility of In-Situ Plasma Cleaning in Normal-conducting Copper Cavities}
% Force line breaks with \\
\author{Qianxu Xia}
\affiliation{Department of Engineering Physics, Tsinghua University, Beijing 100084, People's Republic of China}

\author{Lianmin Zheng}
\affiliation{Department of Engineering Physics, Tsinghua University, Beijing 100084, People's Republic of China}
\affiliation{Key Laboratory of Particle and Radiation Imaging, Tsinghua University, Ministry of Education, Beijing 100084, People's Republic of China}

\author{Yingchao Du}
\affiliation{Department of Engineering Physics, Tsinghua University, Beijing 100084, People's Republic of China}
\affiliation{Key Laboratory of Particle and Radiation Imaging, Tsinghua University, Ministry of Education, Beijing 100084, People's Republic of China}
\homepage{https://www.ep.tsinghua.edu.cn/~Charlie.Author.}
\email{dych@mail.tsinghua.edu.cn}

\date{}

\date{\today}% It is always \today, today,
             %  but any date may be explicitly specified

\begin{abstract}
To assess the feasibility of in-situ plasma cleaning for copper cavities, a 13.56 MHz inductively coupled plasma platform with a built-in coil was developed at Tsinghua University. Experiments were conducted using this platform to optimize plasma discharge parameters and procedures specific to copper cavities. The results indicate that the "Ar/O + Ar/H method" significantly enhances the work function of the copper surface while reducing field enhancement effects induced by surface burrs. Consequently, this study confirms that in-situ plasma cleaning effectively mitigates field emission within copper cavities, thereby enhancing the stability and acceleration gradient of the accelerator system.
\end{abstract}

\maketitle

%\\begin{quotation}The ``lead paragraph'' is encapsulated with the \LaTeX\ \verb+quotation+ environment and is formatted as a single paragraph before the first section heading. (The \verb+quotation+ environment reverts to its usual meaning after the first sectioning command.) Note that numbered references are allowed in the lead paragraph.
%The lead paragraph will only be found in an article being prepared for the journal \textit{Chaos}.\end{quotation}

\section{\label{sec:level1}Introduction}
Particle accelerators are classified into normal-conducting (NC) accelerators and superconducting radio frequency (SRF) accelerators. NC copper cavities face challenges from dark current and electron multipacting. These issues are more pronounced during long pulses \cite{lang1971theory, singh2016raman} or continuous wave (CW) operation. Dark current degrades beam quality, increases thermal load, and elevates radiation levels. It also deteriorates the vacuum environment. Electron multipacting hampers the establishment of high-frequency fields and limits the increase of acceleration gradients in copper acceleration cavities. Both dark current and electron multipacting are manifestations of field emission from the accelerator cavity surface.

The deterioration of field emission is primarily due to two factors. (a)Hydrocarbon organic compounds \cite{knoblock1999field, skriver1992surface} and copper oxide \cite{de2005metal}  (b)Burrs on copper cavity surface \cite{cahill2018high, hu2016influence, bojko2000influence}.

Installing and tuning NC  cavities is complex. Additionally, copper surfaces rapidly adsorb low-boiling-point organic compounds \cite{ciovati2016superconducting, padamsee201750} and react with oxygen to form copper oxides when exposed to the atmosphere, reducing cleaning effectiveness. Thus, it is crucial to keep NC cavities in situ and isolated from the atmosphere during the cleaning process.

Tsinghua University plans to explore the feasibility of in-situ plasma cleaning for NC cavities and has designed a 13.56 MHz inductively coupled (ICP) plasma platform with a built-in coil for this purpose. The platform is used to investigate the reaction principles, and through a combination of simulations and sample experiments, determine the optimal plasma cleaning parameters.

While there are many experiments on plasma cleaning of copper samples \cite{yi1999effects}, research on cleaning  NC cavities is limited \cite{chang2013comparison}. High-frequency copper accelerating structures require a cleaner surface. This study employs an "Ar/O + Ar/H method." It leverages the chemical effects of plasma. The study introduces, for the first time, the oxidation and reduction capabilities of various plasma components. It provides guidance on discharge parameters.

Additionally, NC copper cavities demand extremely low surface roughness. During cleaning, it is essential to passivate burrs without damaging the surface. This study investigates the physical interaction processes between plasma and copper. It elucidates the principles of burr passivation. 

This paper primarily introduces the design of the plasma discharge platform and the interaction between plasma and copper surfaces. The experimental design and result analysis are also discussed. The explored plasma discharge parameters and procedures provide valuable guidance for in-situ plasma cleaning of high-frequency copper cavities.
\section{\label{sec:level1}Cleaning Objectives and Evaluation Methods}
To minimize field emission, it is essential to identify the key parameters that influence it. According to the Fowler-Nordheim equation \cite{fowler1928electron}, the current due to field-induced emission in a high-frequency electric field \cite{padamsee2008rf} can be expressed as follows:

\begin{equation}
I(E) = \frac{A_{FN}A_{e}(\beta_{FN}E)^2}{\phi} \exp\left(-\frac{B_{FN}\phi^{3/2}}{\beta_{FN}E}\right)
\end{equation}

where $I(E)$ represents the field emission current, $E$ denotes the electric field strength at the surface, $A_{e}$ represents the effective emission area, and $A_{FN}$ and $B_{FN}$ are constants (approximately $1.54\times10^6$ and $6.83\times10^3$, respectively). $\phi$ represents the work function, and $\beta_{FN}$ denotes the enhancement factor. To mitigate field-induced emission current, it is necessary to increase the surface work function and decrease the field enhancement factor.

The work function $\phi$ primarily depends on the composition of the copper surface. The work function of copper is 4.65 eV \cite{lang1971theory}, while that of organic substances ranges from 3 eV to 4.15 eV \cite{zhu2023insight}.  Therefore, to enhance the surface work function, it is essential to remove organic substances from the copper surface.

Copper oxides, influenced by surface states, defect states, and surface structure, are susceptible to electron emission\cite{ skriver1992surface} under high electric fields. In copper-based photocathode electron guns, the presence of copper oxides on the cathode surface can destabilize quantum efficiency. To preserve the microwave electric field distribution and maintain the high-quality factor in high-frequency copper cavities, the removal of copper oxides from the cavity surface is crucial.

In the experiment, directly measuring changes in the work function does not clearly indicate which specific conditions have improved. The work function is dependent on the composition of the cavity surface. X-ray Photoelectron Spectroscopy (XPS) is required to determine the elemental composition of the cavity surface. XPS will be used to examine the elemental composition of the copper surface in the 10 nm range. By analyzing changes in the elemental composition of carbon, oxygen, and copper, it is possible to determine trends in the work function of the copper surface. The detection space for XPS is very limited, and the dimensions need to be less than 1 cm \cite{huang2022mechanism,gengenbach2021practical} in all directions. Therefore, the sample is configured as a cylinder composed of oxygen-free copper, with a diameter of 6 mm and a thickness of 3 mm.

The field enhancement factor $\beta_{FN}$ is directly proportional to the surface's sharpness; thus, smoothing the cavity surface is essential.

A laser scanning confocal microscope (LSCM)\cite{yildiz2011imaging,tata1998confocal}will be used to study the morphological changes on the sample surfaces before and after plasma treatment. This technique can scan point-by-point to generate high-resolution images and provide depth information through three-dimensional reconstruction.

The data obtained from LSCM can be used to assess the overall sharpness \cite{hansson2011skewness}of the sample surface using two statistical integral quantities: surface kurtosis (Sku) and surface skewness (Ssk).

\begin{equation} 
    \text{Sku} = \frac{1}{\text{sq}^4}\left[\frac{1}{A} \iint_A Z^4(x, y) \,dx\,dy\right]
\end{equation}
The parameter sq denotes the root-mean-square roughness, and A represents the area. The function z(x,y) describes the height at the location (x,y). A higher Sku value corresponds to a sharper surface, leading to increased field emission severity.
The formula for calculating Ssk is as follows:
\begin{equation} 
    \text{Ssk} = \frac{1}{\text{sq}^3}\left[\frac{1}{A} \iint_A Z^3(x, y) \,dx\,dy\right]
\end{equation}
Ssk describes the deviation of the surface contour line from the reference line (usually the median line). A large Ssk indicates a wide fluctuation range of the surface contour line, with more pronounced surface protrusions and depressions. High Ssk also leads to enhanced field emission.

The objectives of plasma cleaning NC copper cavities are twofold: (a) to remove organic substances and copper oxides from the cavity surface, and (b) to decrease the  Sku and Ssk values.

\section{\label{sec:level1}Model description}

\subsection{\label{sec:level2}The reaction between oxygen plasma and hydrocarbons}

In the microwave electric field, electrons undergo inelastic collisions with molecules or atoms, resulting in the excitation and ionization of Ar/O. Here, considering the plasma chemical effects, only the excitation and ionization \cite{Gao2015Characterization} of oxygen are discussed.

\begin{equation}
\begin{aligned}
    &e + O_2 \rightarrow O + O + e \\
    &e + O_2 \rightarrow OO^* + e \\
    &e + O \rightarrow O^* + e \\
    &e + O_2 \rightarrow O_2^+ + 2e \\
    &e + O \rightarrow O^+ + 2e
\end{aligned}
\end{equation}

Considering the lifetime of excited states, only primary excitation is considered, with excitation energies ranging from approximately 4eV to 6eV. Considering the relatively abundant components in the plasma, their excitation equations \cite{zhu2023insight,lee2006particle}are as follows: 

\begin{equation} 
    O_2\xrightarrow{Ionization   /excitation}O_2 / O/ O^ -/ O _ 2^+/O ^ + /OO ^ * /O^ * 
\end{equation}

Firstly, in oxygen atoms, a lone pair of electrons makes them more prone to attracting additional electrons. As a result, they exhibit stronger oxidizing properties. In contrast, in $O_2$, two oxygen atoms share a pair of electrons. This sharing reduces the electronegativity and electron affinity of each oxygen atom. Thus, the oxidizing properties of single oxygen atoms (O) are much greater than those of $O_2$. Excited-state oxygen atoms, $O^*$, are more active and have higher energy compared to ground-state oxygen atoms. This makes them more likely to participate in chemical reactions. Therefore, the oxidizing properties of $O^*$ are greater than those of O. Oxygen ions, $O^+$, in the ionized state have already lost some electrons. This increases their electron-gaining ability compared to O. Hence, $O^+$ have greater oxidizing properties than O. The majority of $O^*$ are in the primary excitation state with excitation energies below 10 eV.

For the inductively coupled plasma (ICP)\cite{lieberman1994principles}, the sheath voltage typically ranges from 10 to 30 electron volts. Therefore, $O^+$ exhibits greater oxidation ability than $O^*$. The ranking of oxidation ability for each component is determined as follows:
\begin{equation} 
    O ^ +>O_2^+>O^ * >OO^ *>O\gg O_2
\end{equation}

When cleaning oxygen-free copper surfaces, it is crucial to maximize the excitation level of the plasma to increase the reaction rate and extent.

The reaction schematic illustrates that oxygen atoms, excited-state oxygen atoms, and oxygen ions interact with organic substances on oxygen-free copper surfaces, producing $CO_2$ and $H_2O$ at room temperature. The negative Gibbs free energy of this reaction confirms its spontaneity.

In practical applications of plasma cleaning and on experimental platforms, the working gas is continuously cycled through the cavity. This circulation facilitates the removal of $CO_2$ and $H_2O$, which are byproducts of the reaction, and replenishes $O_2$. According to the Arrhenius theorem, this setup ensures a constant reaction rate and high efficiency.

Given that the generated $H_2O$ and decomposed short-chain organic substances may adhere to the cavity surface, baking the cavity during the plasma cleaning process is necessary. This step aims to further enhance the reaction's extent by removing any residual compounds.

\subsection{\label{sec:level2}The reaction between hydrogen plasma and copper oxide}
Ammonia and hydrogen, as reducing gases, can be used to reduce copper oxides. However, the reaction of ammonia plasma with copper leads to the formation of surface protrusions and hard-to-decompose copper nitride. Therefore, Ar/H plasma is chosen as the reducing gas. Considering only the chemical effects, the ionization \cite{2010Low}and excitation process of hydrogen is as follows:

\begin{equation}
\begin{aligned}
    &e + H_2 \rightarrow H + H + e \\
    &e + H \rightarrow H^* + e \\
    &e + H + H_2 \rightarrow H_3^+ + 2e \\
    &e + H_2 \rightarrow H_2^+ + 2e \\
    &e + H \rightarrow H^+ + 2e
\end{aligned}
\end{equation}

Considering the relatively abundant components in the hydrogen plasma, which can undergo excitation and ionization \cite{laidani2004argon,bogaerts2000effects}as follows:
\begin{equation} 
 H_2\xrightarrow{Ionization /excitation}H_2 / H/H ^ + /H_2^+/H_3^+ /H^ * 
\end{equation}
Based on the previous analysis of oxygen plasma, active hydrogen groups show distinct characteristics. H exhibits greater reducing ability than $H_2$. $H^*$ is more active than H. It is more likely to lose electrons. As a result, it is more prone to transferring electrons to other substances. The reducing ability of $H^*$ is greater than that of H.
$H^+$ does not exhibit reducing properties. However, when $H^+$ is accelerated by the sheath voltage, it reaches the surface of the metal. There, $H^*$ rapidly undergoes recombination reactions.
\begin{equation}
 H^++e=H
 \end{equation}
Therefore, when $H^+$ reaches the grounded copper cavity surface, it gains electrons and carries energy. Based on prior analyses of oxygen plasma, the sequence of reducing abilities within hydrogen active groups is tentatively established as follows:
\begin{equation}
   H^+>H_2^+>H_3^+ >H^ *>H\gg H_2
 \end{equation}

 This conclusion is also supported by relevant literature\cite{2014Reduction}. Therefore, efforts should be made to maximize the reduction rate of copper oxide by increasing the degree of ionization.

 It should be noted that active hydrogen groups, which possess superior reducing capabilities, can spontaneously react with oxidized copper at room temperature \cite{Jan0HYDROGEN}. Nonetheless, baking remains a necessary step during plasma treatment. While baking does not substantially increase the reaction rate, this preventive action inhibits the adherence of formed water to the surface. Consequently, this measure prevents impediments to reaction progress and augments the reaction's extent.

\subsection{\label{sec:level2}The reaction between $Ar^+$ and burrs}
The field enhancement factor is closely related to the surface morphology \cite{lieberman1994principles}of the cavity, especially the sharpness of surface spikes. To mitigate field-induced electron emission, it is crucial to reduce the field enhancement factor, which necessitates decreasing surface steepness and the number of spikes. Due to differing acceleration gradients, compared to SRF cavities, NC cavities exhibit higher sensitivity \cite{cahill2018high,wang1997field}to surface spikes. Therefore, passivated burrs play a significant role in improving NC accelerator performance.

\begin{figure}[h]
	\center
	\includegraphics*[width=7cm]{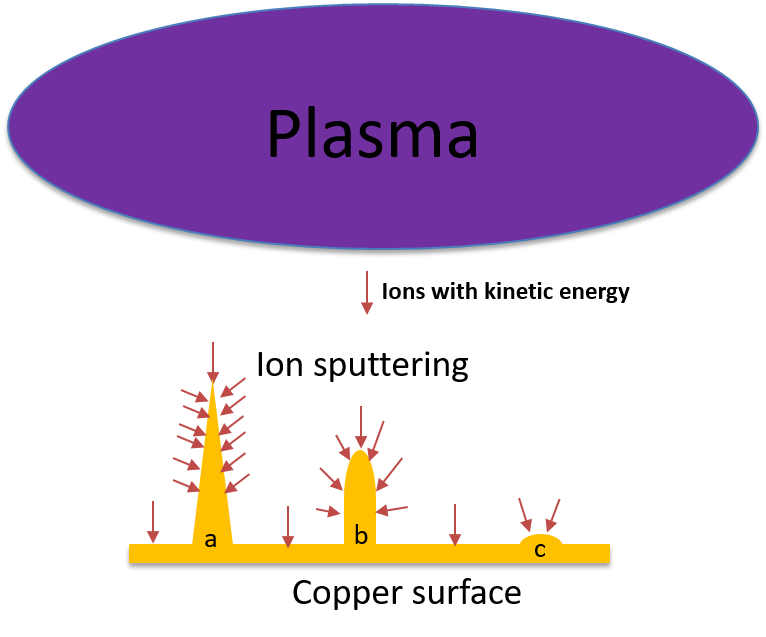}
	\centering
	\caption{Schematic diagram of the plasma passivation burr principle. ``abc'' respectively represent the states of the spikes before plasma treatment, during plasma treatment, and after plasma treatment.}

        \label{fig:burr}
\end{figure}

In the plasma cleaning process, oxygen or hydrogen constitutes only a small fraction of the gas mixture, with argon being the majority. Accelerated by the sheath voltage, $Ar^+$ ions carry kinetic energy and sputter and bombard the surface of oxygen-free copper. As shown in Figure  \ref{fig:burr}, in the plasma environment, since the cavity is grounded, the tips "a" and "b" and the flat surface "c" are at the same potential. However, the electric field distribution induces a "focusing" effect, directing a greater number of $Ar^+$ions to bombard the tips rather than the flat surfaces. This process is analogous to high-power conditioning in accelerator cavities; however, in this case, the heavier ions, rather than electrons, impact the tips, leading to more pronounced morphological changes.

The base area of the tip "a" is very small, resulting in poor thermal conductivity. When subjected to ion bombardment, heat accumulates and the temperature rises until it melts. The flat region "c" experiences significantly less ion bombardment and thus shows no significant changes before and after the treatment. This process improves the surface of the copper.

Due to the characteristics of the plasma sheath electric field, $Ar^+$ ions typically impinge on the spikes at certain angles. This results in a higher sputtering rate on the spikes compared to other flat regions.
The formula \cite{lieberman1994principles}for calculating the sputtering threshold energy ($\zeta_{thr}$)of copper is provided
\begin{equation} 
    \zeta_{thr} \approx8\zeta_{t}(M_\mathrm{i}/M_\mathrm{t})^{2/5}
\end{equation}

$\zeta_{t}$ represents the surface binding energy, approximately equal to the evaporation enthalpy, which can be found in tables, with copper's evaporation enthalpy being approximately 3.5315 eV. $M_i$ is the mass of the incident ions. $M_t$ is the atomic mass of the target. 
The calculations indicate that the sputtering threshold  for copper is approximately 23.44916eV

Here,$\overline{Z_t}$is defined for calculating the sputtering yield
\begin{equation} 
   \overline{Z_t}= \frac{2\operatorname{Z_t}}{\left(\operatorname{Z_i}/{Z_t}\right)^{2/3}+\left(\operatorname{Z_t}/\operatorname{Z_i}\right)^{2/3}} 
\end{equation}
 ${Z_i}/{Z_t} $represents the ratio of the ion number to the target atom number. 

The formula for calculating the sputtering yield $\gamma_{sput}$is as follows:
\begin{equation} 
 \gamma_{sput}\approx\frac{0.06}{\zeta_t}\sqrt{\overline{Z}_t}(\sqrt{\zeta_{i}}-\sqrt{\zeta_{thr}})
\end{equation}
$ \zeta_{i}$ represents the energy of the incident ions.
In ICP plasma, the energy of positive ions bombarding the cavity wall typically remains below 30 electron volts. Substitution into the relevant equation indicates that the sputtering yield is less than 0.063. Therefore, the primary influencing factor is melting induced by ion bombardment.

\section{\label{sec:level1}Experimental Platform Design}
\begin{figure}[h]
	\center
	\includegraphics*[width=10cm]{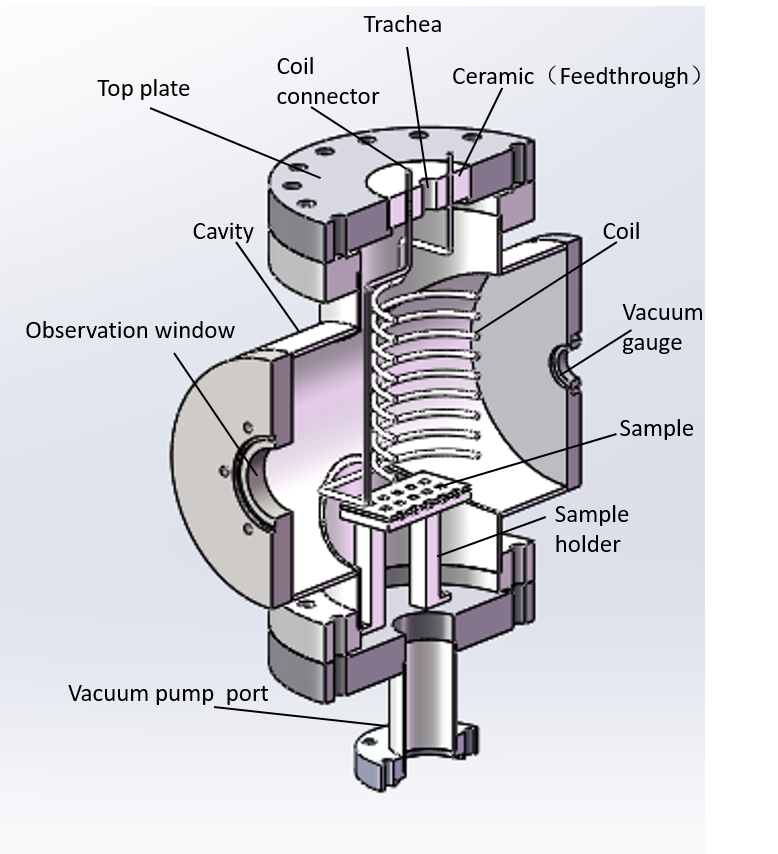}
	\centering
	\caption{Schematic diagram of plasma passivation burr principle.}
        \label{Fig: structure}
\end{figure}

Tsinghua University designed a 13.56 MHz ICP platform with a built-in coil, as shown in Figure\ref{Fig: structure}. The chamber shell is made of 304 stainless steel and the copper coil located inside the chamber, which has been validated in this experiment to better excite the plasma.

 The working gas is introduced into the chamber through a gas pipe located in the top cover and subsequently extracted via an interface in the bottom cover. The vacuum state within the chamber is measured by a thin-film vacuum gauge positioned on the right wall and connected to a vacuum control module. This module actively regulates both an upstream high-speed electric needle valve and a downstream electric baffle valve to sustain the requisite low-pressure environment. The system supports pressure stabilization ranging from 0.1 to 10 Pascals, enabling the assessment of plasma cleaning efficacy under varied pressure conditions.

For spectral observation, a viewing window has been installed on the left wall of the chamber. Spectral analysis, conducted using a spectrometer, facilitates the identification of discharge components and the measurement of both the average plasma temperature and electron density within the chamber.

The microwave system comprises two insulated feedthroughs, a copper coil, an impedance matcher, and a 13.56 MHz microwave source. The copper coil, insulated from the chamber by the feedthrough, connects to the impedance matcher. The sample holder is strategically positioned 3 cm below the copper coil, optimizing plasma density for more effective cleaning. Experimental results indicate that positioning the sample less than 2 cm from the coil during discharge leads to gap breakdown, generating multiple electrical sparks and potentially damaging the sample surface.

The impedance of the plasma fluctuates with variations in power and pressure settings. Tsinghua University has incorporated an impedance matcher capable of continuously adjusting inductance and capacitance to align with the plasma impedance, effectively reducing reflected power to zero.

The copper coil is insulated from the chamber, creating a potential difference due to sheath voltage. A DC isolator is installed to protect the microwave source from potential damage. Additionally, grounding all equipment serves as a safeguard to achieve this protective effect.
\section{\label{sec:level1}Plasma Treatment Process}
 The surface of the accelerator is maintained at a high degree of smoothness and cleanliness to minimize field emission. To accurately simulate real conditions, circular discs will be fabricated using advanced accelerator production techniques. These discs will feature a surface roughness of less than 40 nm. Following fabrication, the discs will be subjected to standard cleaning processes, including hydrogen reduction, ultrasonic water cleaning, and thorough rinsing with pure water.

Eight copper samples were randomly selected and numerically labeled for analysis using XPS and LSCM to document the elemental composition and surface morphology prior to plasma treatment. Six of these samples were directly mounted on the sample holder, while the remaining two were attached via a PTFE base for insulation purposes. The chamber was evacuated for thirty minutes to remove air. Subsequently, a mixture of 1\% oxygen in argon was introduced. The pressure was maintained at 4 pascals, and plasma was excited at a power setting of 300 watts. During this process, heating tapes raised the chamber temperature to 100°C to promote the volatilization of short-chain organic substances and water. This discharge process lasted for thirty minutes. Upon completion, the slide valve was closed to isolate the vacuum pump, and pure argon gas was introduced until the chamber pressure exceeded atmospheric pressure. The vacuum pump was then deactivated, the bottom flange of the chamber was opened, and the samples were rapidly extracted for XPS and LSCM analysis.

The samples were then reintroduced into the chamber, and the initial procedure was replicated to cleanse any organics adsorbed from the air. The supply of Ar/O gas was ceased, and the chamber was evacuated again for thirty minutes to purge residual gases. Subsequently, a mixture containing 3\% hydrogen in argon was introduced. The pressure was set to 10 pascals, and discharge commenced at a power of 100 watts, with the chamber maintained at 100°C. An hour post-discharge, the procedure was reiterated to remove the samples for testing, completing the "Ar/O + Ar/H method."

This methodology, derived from plasma discharge simulations and extensive experimentation, achieves a high cleaning rate and significant cleanliness. Previous tests revealed that using pure argon plasma is ineffectual for cleaning copper surfaces, which is why this approach is not elaborated upon further.

\section{\label{sec:level1}Data analysis and discussion}
\subsection{\label{sec:level2}Element changes before and after plasma treatment}
\begin{figure}[h]
\centering  %图片全局居中
  \subfloat[\label{FIG:cu1}]{
  \includegraphics[width=0.18\textwidth]{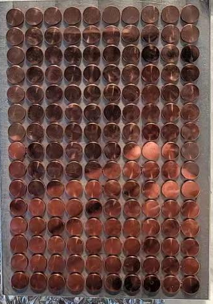}
}
  \subfloat[\label{FIG:cu2}]{

  \includegraphics[width=0.27\textwidth]{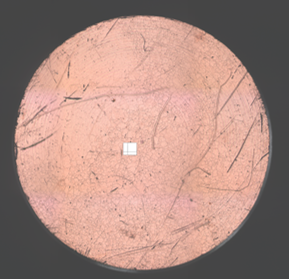}
}
  \caption{(a) Polished and preliminarily cleaned oxygen-free copper sample (b) Copper sample surface magnified 50 times}
    
\end{figure}

Figure \ref{FIG:cu1}  presents a microscopic pre-treatment image of an oxygen-free copper sample, identified as number 3. The surface of this sample is smooth and lustrous, resembling a mirror. Additionally, Figure \ref{FIG:cu2} includes a 50x magnified view of the same sample's surface, which displays a brownish-yellow coloration with numerous dull patches, indicative of the presence of organic residues and copper oxides.

XPS analysis of sample 3 has determined its surface elemental composition as follows: carbon (C) constitutes 60.8\%, oxygen (O) 25.8\%, copper (Cu) 11.7\%, sulfur (S) 1.6\%, and nitrogen (N) 0.1\%.

In Sample 3, carbon and oxygen elements are derived from volatile organic compounds (VOCs) and copper oxides, respectively. Sulfur originates from the cutting oil used during processing, while nitrogen is attributable to atmospheric adsorption. Test results reveal that the copper content on the surfaces of samples produced by accelerator manufacturing processes is significantly diminished due to air pollution and oxidation. Thus, it is imperative to prevent secondary contamination from the air. This finding highlights the critical role of in-situ plasma cleaning in maintaining the integrity of NC cavities.

The surface elemental compositions of eight copper samples were assessed using XPS after they underwent two rounds of plasma treatment. Samples 7 and 8, isolated on pedestals, exhibited negligible changes in their elemental compositions pre- and post-treatment. Unlike the grounded samples, these encountered reduced ion bombardment due to charge accumulation, leading to fewer elemental alterations. In contrast, the other six samples, which were in direct contact with the base, showed more pronounced changes, demonstrating the heightened chemical reactivity of ions within the plasma. Consequently, the sequence of chemical activity is established as $O_2^+ > O^+$ and $H_3^+ > H^+$.

Given the consistent elemental changes observed in the six samples that were in direct contact with the base, Sample 3 was selected for further detailed analysis.

\begin{figure}[h]
%\centering  %图片全局居中

\subfloat[\label{FIG:CUO}]{
\includegraphics[width=0.22\textwidth]{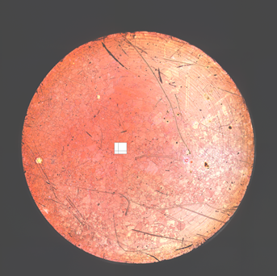}
}
\subfloat[\label{FIG:CUH}]{
\includegraphics[width=0.24\textwidth]{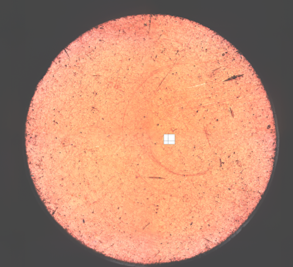}
}
\caption{(a) Oxygen-free copper surface after Ar/O plasma  treatment (b) Oxygen-free copper surface after "Ar/O +Ar/H" plasma  treatment}
\end{figure}

Figure \ref{FIG:CUO} illustrates the morphology of Sample 3 post Ar/O plasma treatment. The previously dull areas have diminished, and the surface organic content has been significantly reduced. Parts of the surface exhibit a reddish-brown hue, indicative of cuprous oxide formation. Subsequently, Figure\ref{FIG:CUH} depicts Sample 3 following the "Ar/O + Ar/H" plasma treatment, where the surface appears orange-yellow with almost no dull areas. This transformation suggests a substantial reduction in surface organics and copper oxide.

Table \ref{tab:Surface} details the elemental changes of Sample 3 before and after plasma treatment.

The data presented in the table illustrate significant changes in elemental composition following plasma treatment. After treatment with Ar/O plasma, there is a notable reduction in carbon content, signifying the effective removal of organic matter. Concurrently, an increase in oxygen content reflects the oxidation of the copper surface. Additionally, there are slight increases in both nitrogen and copper content.

Following the "Ar/O + Ar/H method" plasma treatment, there is a slight increase in nitrogen content, while both oxygen and carbon content decrease significantly, indicating a substantial reduction of organic matter and copper oxides. Sulfur content diminishes to undetectable levels, and copper content nearly doubles, compared to pre-treatment levels.

\begin{table}
\caption{\label{tab:Surface}Surface element ratio of oxygen-free copper samples before and after plasma treatment}
\begin{ruledtabular}
\begin{tabular}{cccc}
\textbf{Element} & \textbf{Initial} & \textbf{O/Ar} & \textbf{O/Ar+H/Ar} \\ \hline 
C & 60.8\% & 36.5\% & 44.3\% \\ 
O & 25.8\% & 41.4\% & 20.5\% \\  
Cu & 11.7\% & 18.1\% & 28.3\% \\ 
N & 0.1\% & 2.1\% & 3.2\% \\ 
S & 1.6\% & 1.9\% & - \\
\end{tabular}
\end{ruledtabular}

\end{table}

The  XPS spectrum of Sample 3 is presented below, providing a clearer depiction of the changes in elemental composition before and after treatment.
\begin{figure}[h]
\centering
\begin{minipage}{0.45\textwidth}
\centering
\subfloat[\label{fig:XPSa}]{
\includegraphics[width=\linewidth]{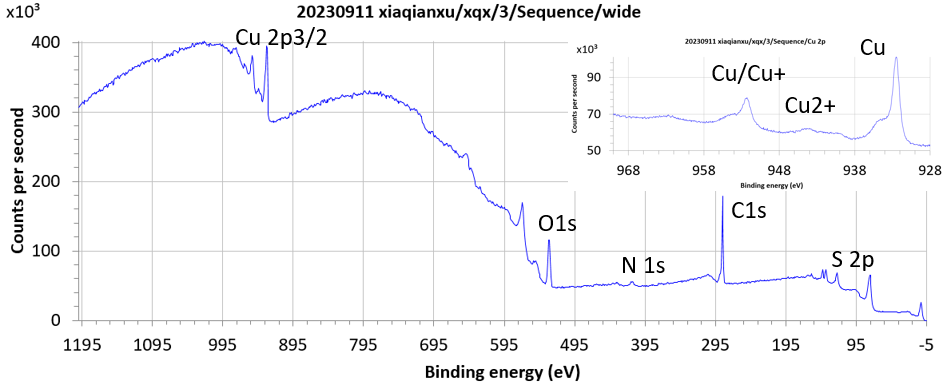}  }
\end{minipage}
\begin{minipage}{0.45\textwidth}
\centering
\subfloat[\label{fig:XPSb}]{
\includegraphics[width=\linewidth]{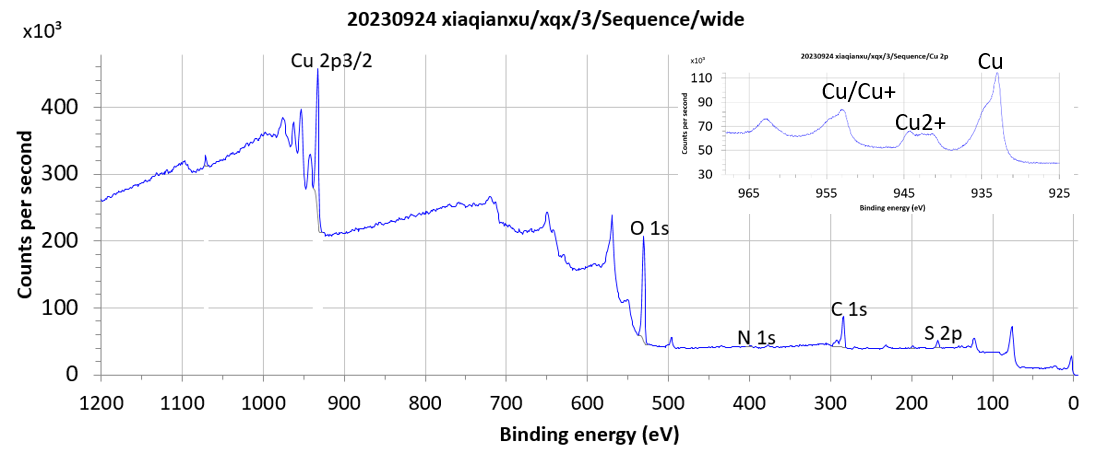}  }
\end{minipage}
\begin{minipage}{0.45\textwidth}
\centering
\subfloat[\label{fig:XPSc}]{
    \includegraphics[width=\linewidth]{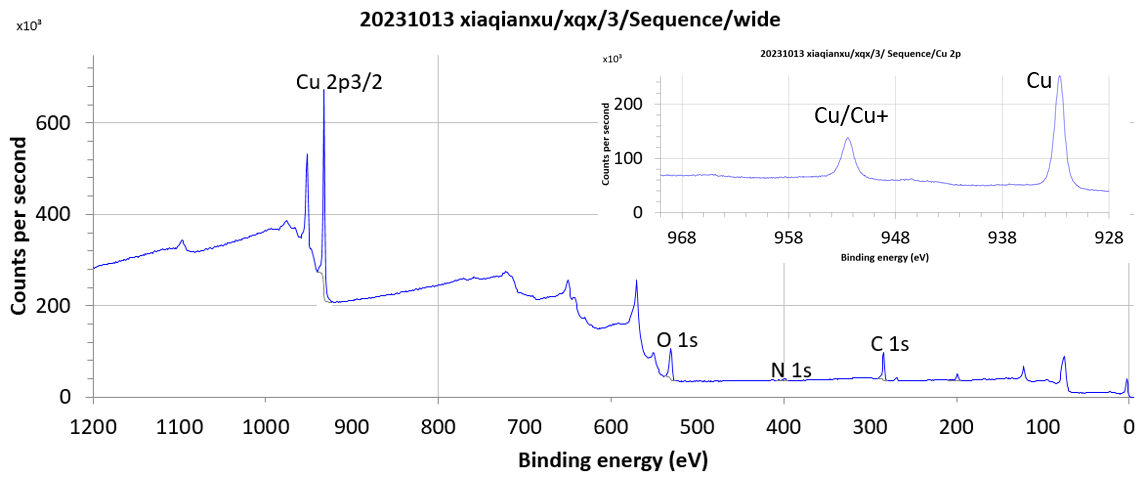}  }
\end{minipage}
\caption{XPS Full Spectrum and High-Resolution Spectrum of Copper on the Surface of Oxygen-Free Copper Before and After Plasma Treatment: (a) Before Plasma Treatment, (b) Ar/O Plasma Treatment, (c) "Ar/O +Ar/H" Plasma Treatment.}
\label{fig:XPS}
\end{figure}

Figure \ref{fig:XPS} shows the XPS spectra of Sample 3 before and after plasma treatment, further validating the effectiveness of plasma cleaning.

Figure \ref{fig:XPSa} displays the spectrum of the sample before cleaning, where the copper element peak is not prominent. In the high-resolution XPS spectrum, a small amount of cuprous oxide can be observed, and the carbon peak is very distinct.

\begin{figure*}[hbt!]
  \centering
  \begin{minipage}{0.32\textwidth}
    \centering
    \subfloat[\label{FIG:1BF}]{\includegraphics[width=\linewidth]{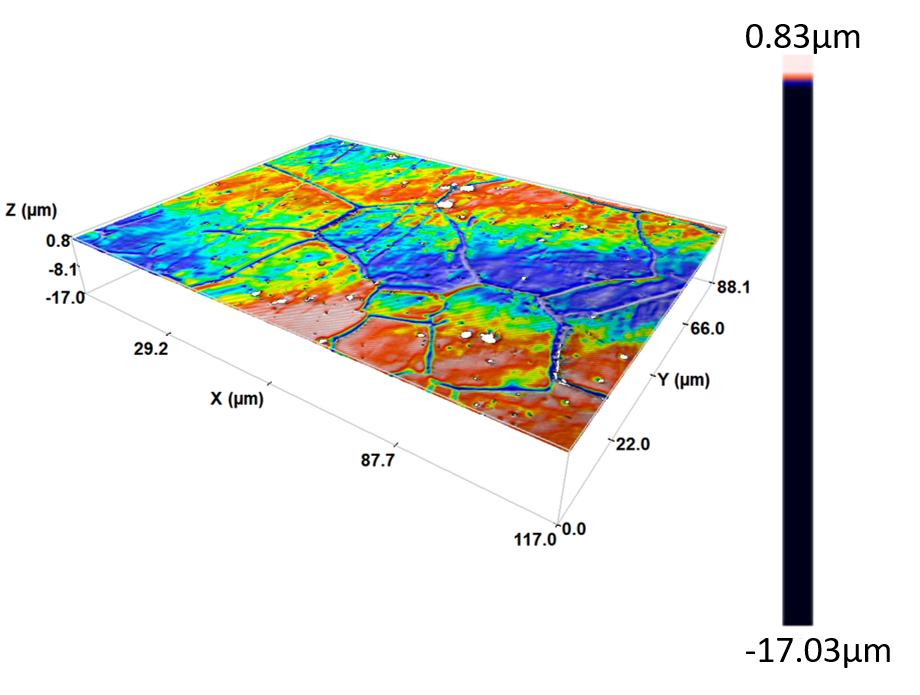}}
  \end{minipage}%
  \begin{minipage}{0.32\textwidth}
    \centering
    \subfloat[\label{FIG:1AFO}]{\includegraphics[width=\linewidth]{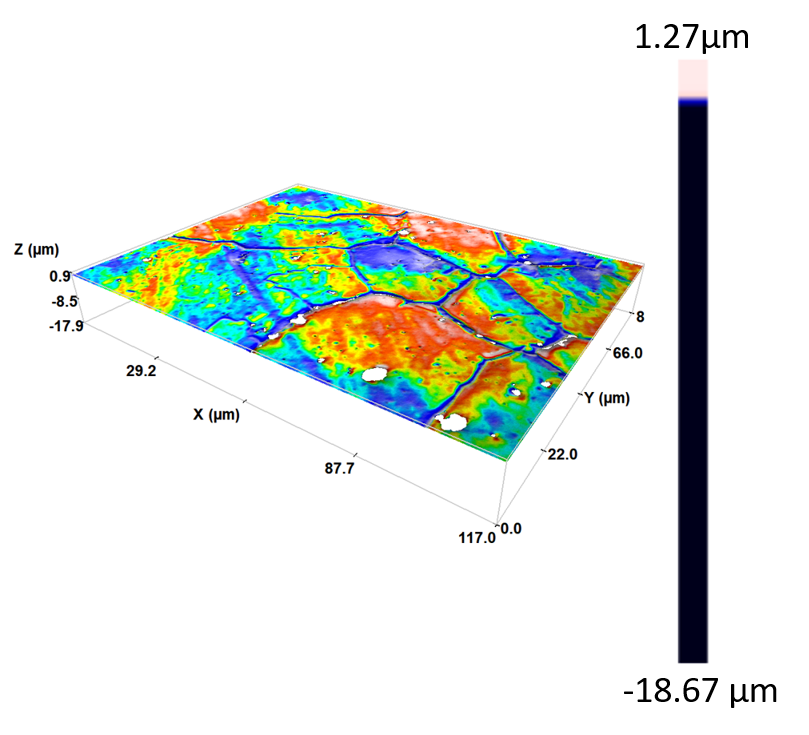}}
  \end{minipage}%
  \begin{minipage}{0.32\textwidth}
    \centering
    \subfloat[\label{FIG:1AFH}]{\includegraphics[width=\linewidth]{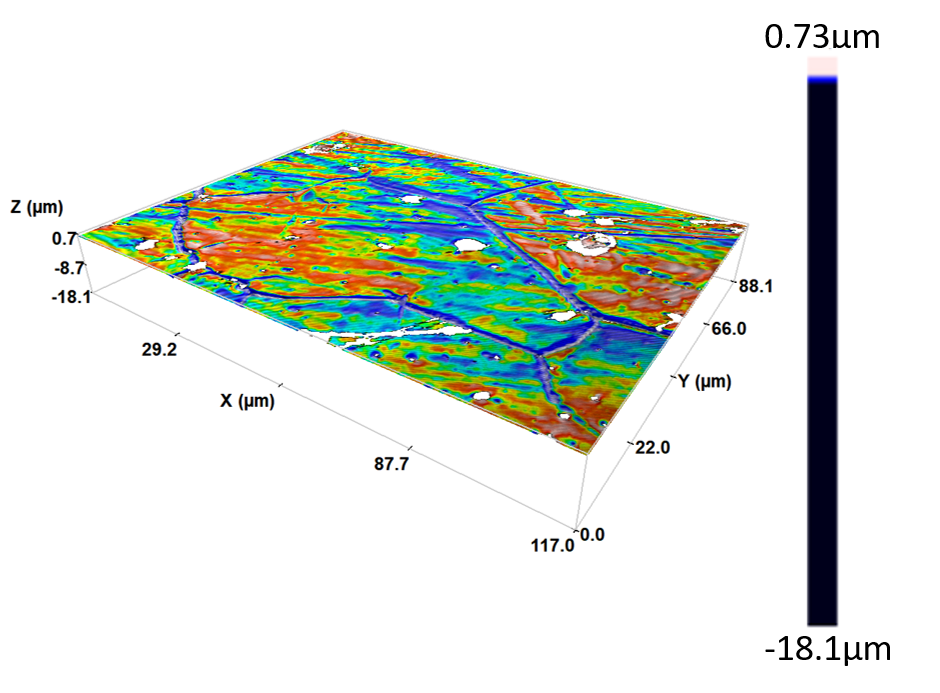}}
  \end{minipage}

  \begin{minipage}{0.32\textwidth}
    \centering
    \subfloat[\label{FIG:2BF}]{\includegraphics[width=\linewidth]{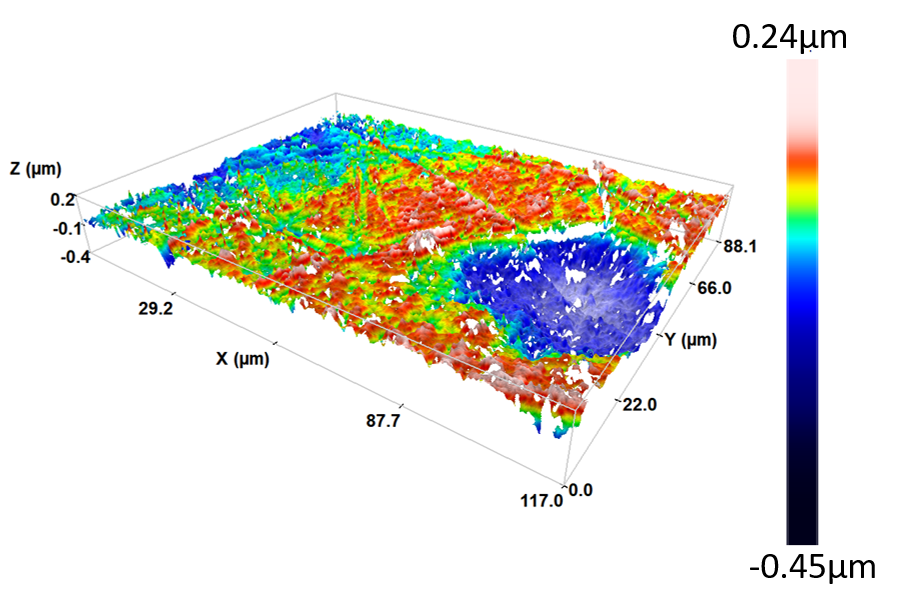}}
  \end{minipage}%
  \begin{minipage}{0.32\textwidth}
    \centering
    \subfloat[\label{FIG:2AFO}]{\includegraphics[width=\linewidth]{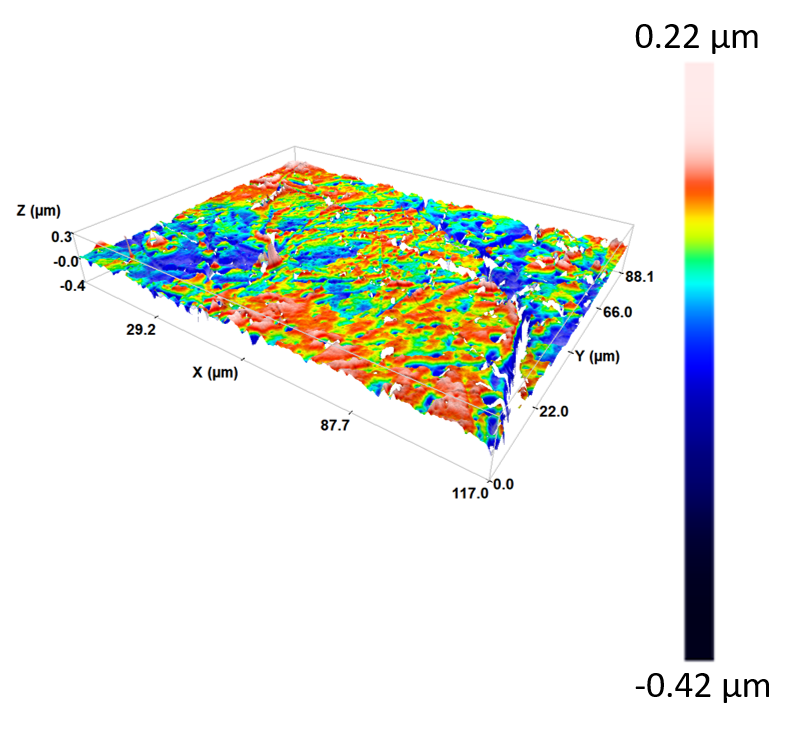}}
  \end{minipage}%
  \begin{minipage}{0.32\textwidth}
    \centering
    \subfloat[\label{FIG:2AFH}]{\includegraphics[width=\linewidth]{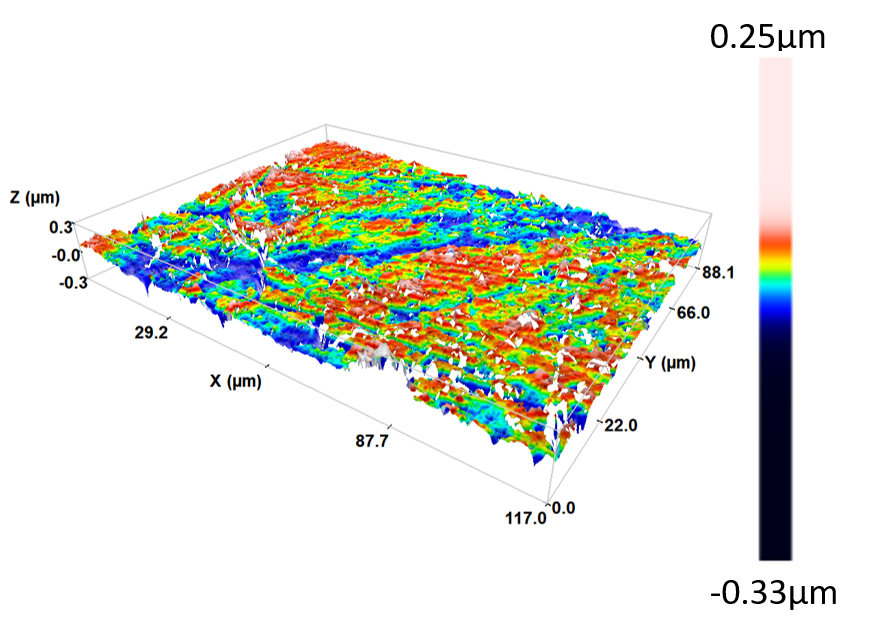}}
  \end{minipage}

  \begin{minipage}{0.32\textwidth}
    \centering
    \subfloat[\label{FIG:3BF}]{\includegraphics[width=\linewidth]{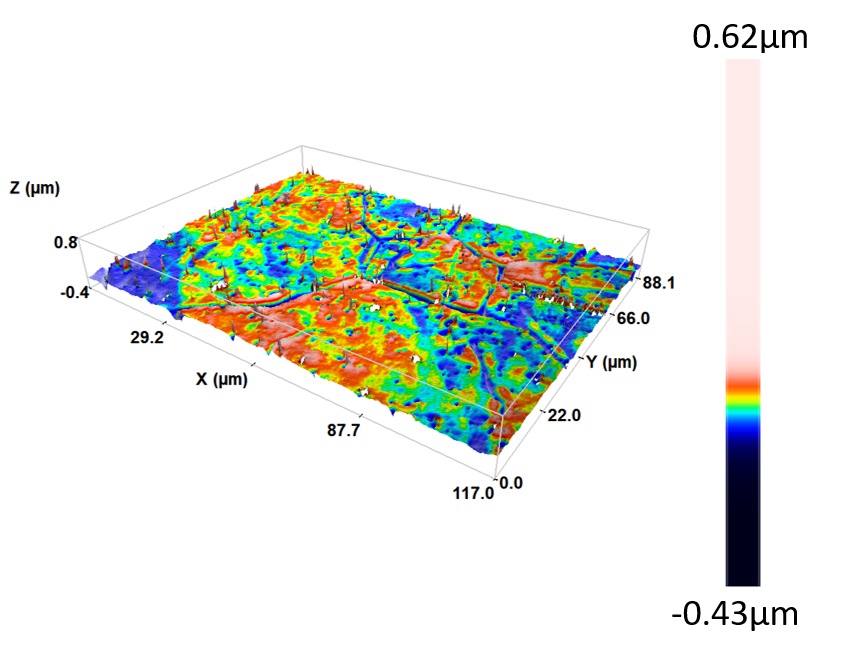}}
  \end{minipage}%
  \begin{minipage}{0.32\textwidth}
    \centering
    \subfloat[\label{FIG:3AFO}]{\includegraphics[width=\linewidth]{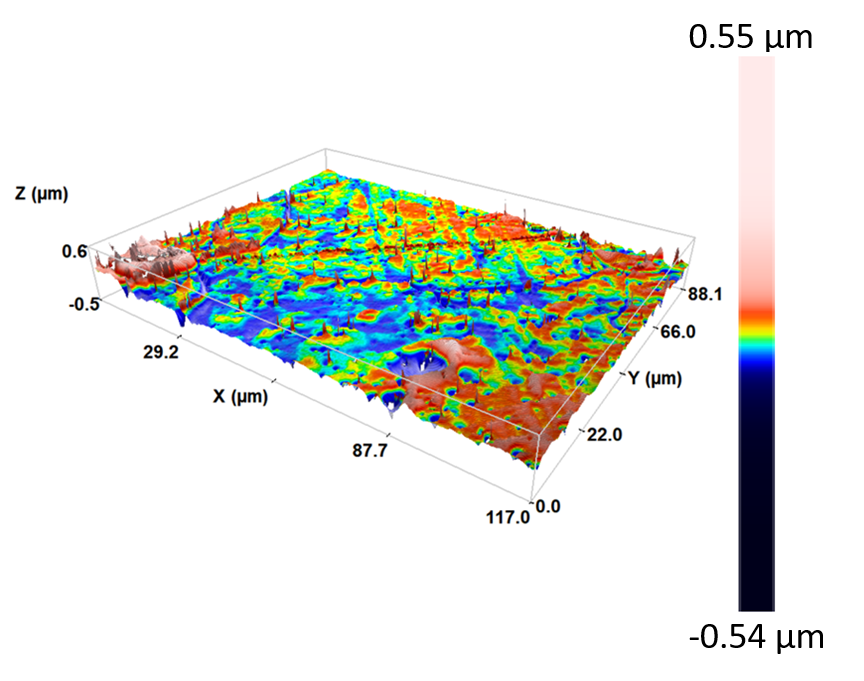}}
  \end{minipage}%
  \begin{minipage}{0.32\textwidth}
    \centering
    \subfloat[\label{FIG:3AFH}]{\includegraphics[width=\linewidth]{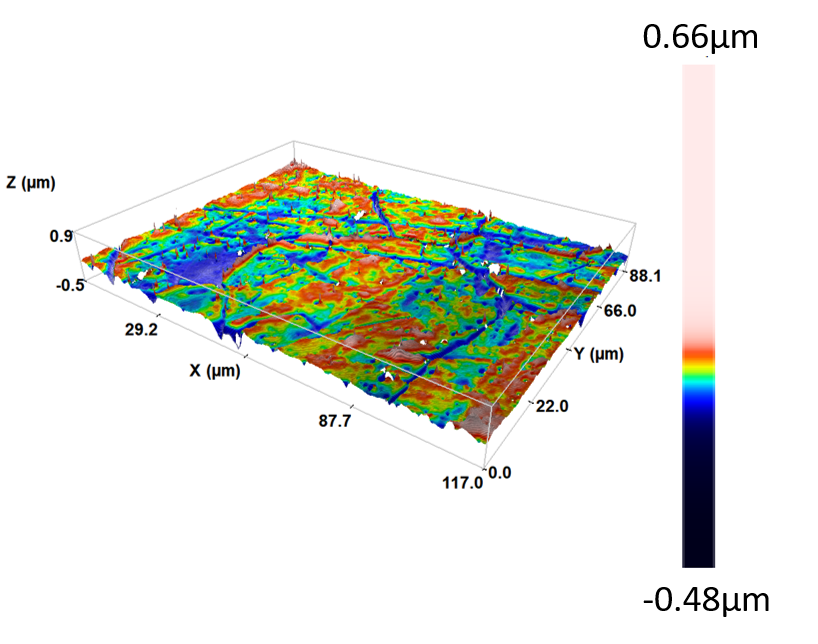}}
  \end{minipage}

  \caption{Confocal Lens Results of Oxygen-Free Copper Surface Before and After Plasma Treatment: (abc) Before Plasma Treatment, (def) Ar/O Plasma Treatment, (ghi) "Ar/O +Ar/H" Plasma Treatment.}
  \label{FIG:BFP}
\end{figure*}

Figure \ref{fig:XPSb} shows the elemental content after Ar/O   plasma treatment. The carbon peak is significantly reduced. The copper and oxygen peaks are significantly increased. In the high-resolution XPS spectrum, copper oxides and cuprous oxide slightly increase.

Figure \ref{fig:XPSc} presents the spectrum after the "Ar/O +Ar/H " plasma treatment. The carbon peak increases compared to after the Ar/O   plasma treatment but remains much lower than before the plasma treatment. The oxygen peak significantly decreases, and the copper peak significantly increases. In the high-resolution XPS spectrum, copper oxides are significantly reduced. This result also proves that this type of plasma treatment is effective in improving the surface work function of copper.

\subsection{\label{sec:level2}Surface morphology changes before and after plasma treatment}
Figure \ref{FIG:BFP}  shows the surface morphology of oxygen-free copper samples before and after plasma treatment, as measured by LSCM. From left to right, the three columns respectively show the surface morphology of Samples 3, 4, and 5 before plasma treatment, after Ar/O   plasma treatment, and after "Ar/O +Ar/H " plasma treatment.

The first column of images shows the samples before treatment, with high peaks and depths and many blank areas. The blank areas indicate that their height or depth is significantly greater \cite{yildiz2011imaging}than the surrounding areas. The second column shows the samples after Ar/O  plasma treatment, with significantly reduced peaks and depths and fewer blank areas. The third column represents the samples after "Ar/O +Ar/H " plasma treatment, with slightly increased peaks and depths but still significantly lower than the untreated samples. Almost no blank areas are present, but the surface has a lot of microscopic fluff on the order of nm.

To thoroughly and accurately evaluate the changes in surface morphology of the samples before and after plasma treatment, particularly to determine the passivation of protrusions, it is essential to measure Sku and Ssk. LSCM was employed to assess the surface conditions in three distinct areas of each sample—the center and the two sides. The measurements revealed consistent trends across these areas both before and after plasma treatment, indicating uniform changes. Consequently, this report focuses on the experimental results from the central region of the samples.

\begin{figure}[!h]
\centering
\begin{minipage}{0.4\textwidth}
\centering
\subfloat[\label{fig:sku}]{
\includegraphics[width=\linewidth]{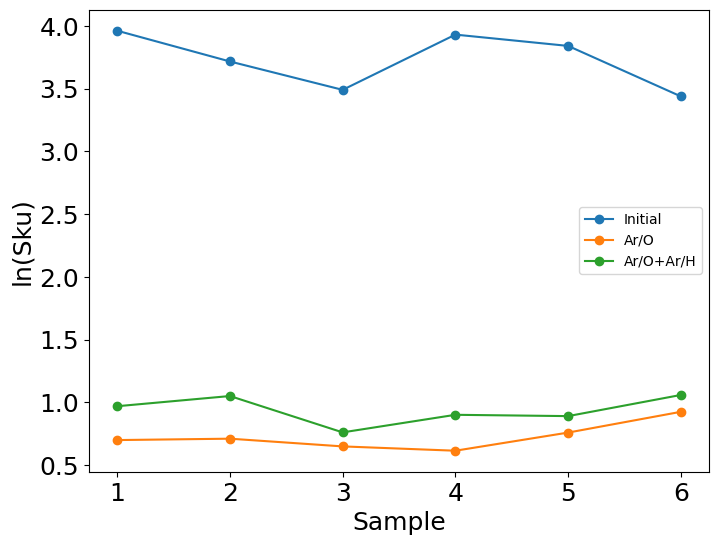}  }
\end{minipage}
\begin{minipage}{0.4\textwidth}
\centering
\subfloat[\label{fig:ssk}]{
\includegraphics[width=\linewidth]{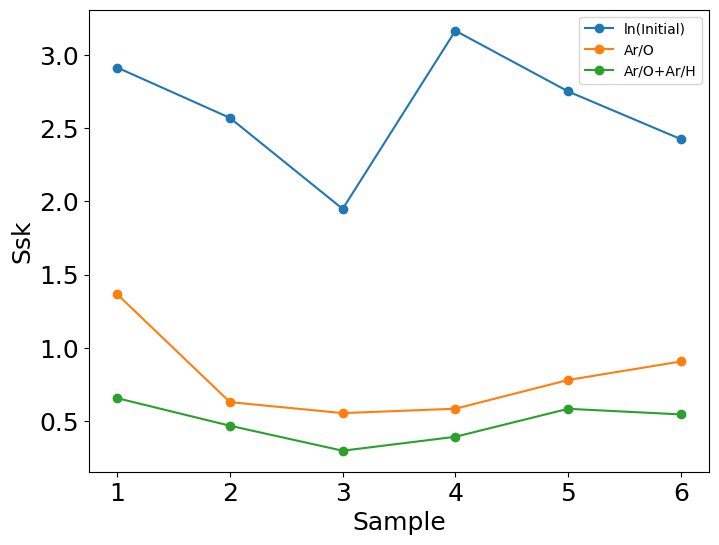}  }
\end{minipage}
\begin{minipage}{0.4\textwidth}
\centering
\subfloat[\label{fig:rough}]{
    \includegraphics[width=\linewidth]{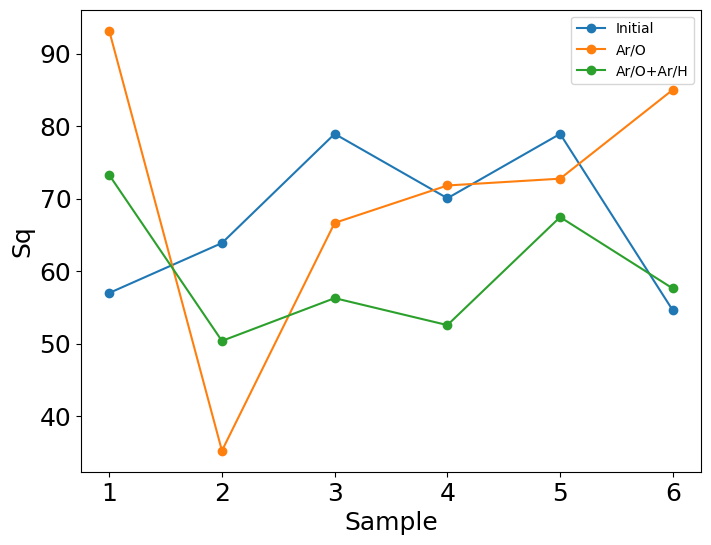}  }
\end{minipage}
\caption{Lens Results of Oxygen-Free Copper Surface Before and After Plasma Treatment: (a) Before Plasma Treatment, (b) Argon-Oxygen Plasma Treatment, (c) Argon-Hydrogen Plasma Treatment.}
\end{figure}

Figure \ref{fig:sku}  shows the changes in Sku before and after plasma treatment, with the vertical axis plotted on a logarithmic scale with base 10. The Sku values of the sample surfaces before treatment are extremely high, significantly decreasing after Ar/O   treatment, indicating reduced surface sharpness and significant passivation of protrusions. After "Ar/O +Ar/H " plasma treatment, the values slightly increase.

Figure \ref{fig:ssk} shows the changes in surface Ssk before and after plasma treatment. The initial Ssk value was large, so the natural logarithm was taken. After Ar/O  plasma treatment, Ssk significantly decreases, indicating that the copper sample surfaces become progressively smoother and the protrusions are effectively flattened. After "Ar/O +Ar/H " plasma treatment, the values further decrease.

Figure\ref{fig:rough} shows the changes in roughness before and after plasma treatment, with no significant changes observed. Samples 2, 3, 4, and 5, located near the center of the coil, show reduced roughness after treatment. This reduction may be attributed to factors such as sputtering angles and ion density distribution, which will be further validated through subsequent simulation calculations and experiments.

\subsection{Discussion}
Due to the clearance between the sample holder and the chamber, achieving the required temperature during baking proves challenging. Furthermore, for safety reasons, the heating tape's coverage during the "Ar/O + Ar/H" plasma treatment is less effective than during the Ar/O plasma treatment, resulting in a slightly less effective removal of organic matter.

The mechanical pump fitted to the device lacks sufficient pumping speed, preventing it from avoiding atmospheric nitrogen contamination during plasma treatment. However, in practical NC cavity plasma cleaning experiments, this issue is mitigated by a superior vacuum environment.

Since the samples must be removed and transported to relevant institutions for XPS testing, exposure to the atmosphere is inevitable. Despite the use of a vacuum box, short-term atmospheric contamination cannot be completely avoided, leading to suboptimal results.

Nevertheless, it has been demonstrated that "Ar/O + Ar/H" plasma treatment effectively reduces the work function of the copper surface, as particularly evidenced by the XPS spectra.

In terms of enhancing surface morphology, "Ar/O + Ar/H" plasma treatment is notably effective. Although it does not substantially reduce roughness, it significantly decreases Sku and Ssk values, passivates surface burrs, and thereby reduces the field enhancement factor.

During the reduction of copper oxides to elemental copper in Ar/H plasma, a unique crystalline orientation phenomenon known as whisker growth\cite{wehner1983whisker} was observed, characterized by the formation of very fine "whisker-like" structures. This phenomenon may be exacerbated\cite{hu2016influence} by the presence of silver impurities.

In subsequent NC cavity experiments, it was demonstrated that whiskers formed by plasma cleaning rapidly vanish during high-power conditioning, confirming that whiskers do not increase the field enhancement factor.

\section{\label{sec:level1}Conclusion}
Tsinghua University has developed a 13.56 MHz ICP platform with an integrated coil to evaluate the feasibility of in-situ plasma cleaning for normal conducting (NC) cavities. The results from this platform indicate that plasma cleaning significantly enhances the surface conditions of copper samples. XPS analyses confirm the plasma's ability to effectively remove organic residues and reduce surface copper oxides, thus substantially improving the work function of the surface. LSCM reveals significant alterations in the Sku and Ssk of the copper surfaces, indicating effective passivation of surface burrs and a reduction in the field enhancement factor. Consequently, this in-situ plasma treatment proves to be an efficient method for reducing field-induced emissions on NC cavity surfaces.

This experimental platform utilizes ultra-clean, ultra-smooth oxygen-free copper samples as proxies for the inner surfaces of normal conducting (NC) cavities to optimize plasma cleaning parameters and processes. Through computational simulations and iterative experiments, a robust cleaning protocol was established. The protocol maintains a constant temperature of 100°C. It employs a mixture of 1\% oxygen in argon, sustaining a pressure of 4 pascals and a microwave power input of 300 watts for 30 minutes. Following a 30-minute exhaust gas venting period, a mixture of 3\% hydrogen in argon is introduced under a pressure of 10 pascals and microwave power of 100 watts for an hour. The experiment elucidated the oxidation potency sequence in oxygen plasma as $O^+ > O_2^+ > O^* > OO^* > O \gg O_2$, and the reduction sequence in hydrogen plasma as $H^+ > H_2^+ > H_3^+ > H^* > H \gg H_2$, contingent on the insulation of copper samples from the 304 chamber.

Considering the complexities of in-situ plasma cleaning for NC cavities, Tsinghua University is committed to continuing its research to enhance and refine the cleaning process, aiming to develop superior methodologies.
\begin{acknowledgments}
This work is supported by the National Natural Science Foundation of China (NSFC) under Grant No. 12275149
\end{acknowledgments}

\section*{ADUTHOR DECLARATIONS}

\subsection*{Conflict of Interest}
The authors have no conflicts to disclose.

\section*{DATA AVAILABILITY}

The data that support the findings of this study are available from the corresponding author upon reasonable request.

\section*{References}
\nocite{*}
\bibliography{aipsamp}% Produces the bibliography via BibTeX.

\end{document}